\documentclass[aps,prd,twocolumn,preprintnumbers,nofootinbib,superscriptaddress,amsmath]{revtex4-2}
\usepackage[english]{babel}
\usepackage{bm}
\usepackage[utf8]{inputenc}
\usepackage[colorinlistoftodos, color=green!40, prependcaption]{todonotes}
\usepackage{amssymb}
\usepackage{graphicx}
\usepackage{natbib}
\usepackage[normalem]{ulem}

\usepackage[pdftex, pdftitle={Article}, pdfauthor={Author}]{hyperref} 

\begin{document}

\title{Thawing Dark Energy and Massive Neutrinos in Light of DESI}

\author{Gabriel Rodrigues}
\email{gabrielrodrigues@on.br}
\author{Rayff de Souza}
\email{rayffsouza@on.br}
\author{Jamerson Rodrigues}
\email{jamersonrodrigues@on.br}
\author{Jailson Alcaniz}
\email{alcaniz@on.br}
\affiliation{Observatório Nacional, Rio de Janeiro - RJ, 20921-400, Brasil}
\date{\today}

\begin{abstract}
Recent analyses have shown that a dynamic dark energy modeled by the CPL parameterization of the dark energy equation of state (EoS) can ease constraints on the total neutrino mass compared to the standard $\Lambda$CDM model. This helps reconcile cosmological and particle physics measurements of $\sum m_\nu$. In this study, we investigate the robustness of this effect by assessing the extent to which the CPL assumption influences the results. We examine how alternative EoS parameterizations - such as Barboza-Alcaniz (BA), Jassal-Bagla-Padmanabhan (JBP), and a physically motivated thawing parameterization that reproduces the behavior of various scalar field models - affect estimates of $\sum m_\nu$. Although both the BA and JBP parameterizations relax the constraints similarly to the CPL model, the JBP parameterization still excludes the inverted neutrino mass hierarchy at $\sim 2.1\;\sigma$ with $\sum m_\nu < 0.096$\;eV. The thawing parameterization excludes the inverted hierarchy at $\sim 3.3\sigma$ and yields tighter constraints, comparable to those of the $\Lambda$CDM model, with $\sum m_\nu < 0.071$\;eV. Finally, we show that the thawing model can be mapped into the BA and JBP $w_0$–$w_a$ parameter space, with the apparent preference for the phantom regime actually supporting quintessence (non-phantom) models. 
\end{abstract}




\maketitle



\section{Introduction}

In the standard cosmology, a dark energy (DE) component drives the current cosmic acceleration. This component possesses negative pressure and is characterized by an Equation of State (EoS) parameter defined as $w= p/\rho$, where $p$ is the dark energy pressure and $\rho$ its energy density. A natural theoretical representation of DE is the vacuum energy density, also known as the cosmological constant $\Lambda$, for which $w = -1$. However, despite being able to explain most of the present-day observational data, the standard $\Lambda$-Cold Dark Matter ($\Lambda$CDM) model faces tensions in measurements of some cosmological parameters. The most prominent of these discrepancies is a tension of $\sim 5\sigma$ between the Hubble Constant $H_0$ measurements in the late- and early-time Universe. Furthermore, $\Lambda$ is associated with  well-known fundamental problems in theoretical physics~\cite{Weinberg:1988cp,Weinberg:2000yb,Padmanabhan:2002ji,Alcaniz:2006ay}.

In the context of General Relativity, two major frameworks have emerged as alternatives to model the Universe's accelerated expansion (see e.g.~\cite{Copeland:1997et,Amendola:1999er,Lima:2000jc,Dev:2002qa,Bagla:2002yn,Sahni:2002dx,Carvalho_2006}): the scalar field approach, exemplified by quintessence models, and phenomenological parameterizations of the dark energy equation of state (EoS), such as the widely used Chevallier–Polarski–Linder (CPL), Barboza–Alcaniz (BA), and Jassal–Bagla–Padmanabhan (JBP) forms~\cite{2001IJMPD..10..213C,2003PhRvL..90i1301L,Jassal_2005,Barboza08,Barboza:2009ks}. In both frameworks, the dark energy EoS is a dynamical quantity, $w(z)$. 

Quintessence models describe dark energy as a slowly rolling scalar field with a canonical kinetic term and a potential $V(\phi)$ that drives its evolution. A key feature of these models is that the EoS remains strictly above the phantom divide, $w(z) > -1$, unless the theory is extended to include exotic components. On the other hand, phenomenological parameterizations do not assume an underlying physical origin for DE but instead provide flexible, model-independent forms for $w(z)$. These can allow the EoS to cross the phantom divide and are useful for fitting cosmological data. However, as discussed in~\cite{Shlivko:2024llw, Wolf:2023uno,Wolf:2024eph, Shlivko:2025fgv,deSouza:2025rhv}, a phantom-like behavior in parameterized EoS models — such as $w(z) < -1$ in the CPL framework — does not necessarily imply the existence of or even preference for phantom dark energy.  When mapping physically motivated quintessence models into the CPL $w0$-$wa$ parameter space, many fall within the so-called phantom region, indicating that such behavior can emerge from standard scalar field dynamics. Therefore, while these parameterizations should not be dismissed — as they are useful in searching for deviations from $\Lambda$
— they are limited in their ability to reveal the physical nature of DE.

To evaluate the observational viability of dynamical dark energy (DDE) models, various probes from both the early and local Universe have been employed~\cite{Wang_2022,peracaula2019,Pan_2019,Escamilla_Rivera_2022,Valentino_2020,Yang_2019,Yang_2020}.  CMB data alone cannot strongly constrain these models due to a geometric degeneracy between late-time parameters and the angular diameter distance to the last scattering surface. Breaking this degeneracy requires the inclusion of local Universe observables.  Although combining CMB with SN data improves constraints, no clear preference for DDE emerged initially. This remained the case until very recently, when BAO observations from the Dark Energy Spectroscopic Instrument (DESI), combined with Planck CMB and Pantheon+ SN data, showed a significant preference for DDE using the CPL EoS parameterization, $w(a) = w_0 + w_a(1-a)$~\cite{DESI:2025zgx}. Even more recently, it has been shown that this preference is not limited to the CPL but extends to other parameterizations as well. This result further reinforces the preference indicated by parametric analyses of the DESI and SNe data for DDE models~\cite{Giare04}.

{Beyond the DDE discussion, the late-time universe is plagued by further uncertainties related to the background and perturbations evolution, and the neutrino physics is of particular interest in this picture. In the standard scenario, outlined by the Standard Model of fundamental particles (SM), neutrinos are massless particles. Such a conclusion is challenged by the number density measurements of the solar and atmospheric neutrinos, pointing to an oscillatory behavior of the three flavors. Oscillation experiments are insensitive to the absolute mass scale of the neutrinos, but useful to probe their squared mass differences. The latter results from atmospheric neutrinos point to $|\Delta m_{32}^2| \approx 2.4 \times 10^{-3}\; \text{eV}^2$, while from solar neutrinos one obtains $|\Delta m_{21}^2| \approx 7.4 \times 10^{-5}\; \text{eV}^2$~\cite{ParticleDataGroup:2024cfk}. This configuration enables at least two non-relativistic species of neutrinos. Lower limits on the sum of their masses can also be estimated:  $\sum m_\nu > 0.06\; \text{eV}$ for the Normal Ordering schemes (NO), while the Inverted Ordering (IO) requires  $\sum m_\nu > 0.10\; \text{eV}$. }

Back to the cosmological perspective, neutrinos have a well-marked role in the  expansion of the cosmos. They act as radiation in the very early universe, later changing to a hot dark matter behavior after the non-relativistic transition. At the perturbative level, the neutrinos' dispersion acts to suppress the growth of perturbations at small scales, characterized by their free-streaming and, ultimately, by the sum of their masses. Increasing $\sum m_\nu$ also changes the amplitude of the distance measurements in late-time Universe. In particular, the discrepancy between the values of $H_0 r_d$ measured by CMB observations and BAO measurements can be mitigated by reducing $\sum m_\nu$ in the context of $\Lambda$CDM \cite{DESI:2025ejh}. One of the current tightest constraints, $\sum m_\nu < 0.07\; \text{eV}$ at 95\% confidence level (C.L), comes from the combination of CMB and DESI BAO data within the $\Lambda$CDM framework, showing a clear preference for the NO~\cite{DESI:2025zgx, Jimenez22} (see also~\cite{Gariazzo:2022ahe}). More recently, the upper limit $\sum m_\nu < 0.042$\;eV was obtained using CMB combined with DESI BAO,  galaxy cluster measurements, and a $H_0$ prior from SH0ES~\cite{Jiang24}. Therefore, the current cosmological limits for the $\Lambda$CDM are on the verge of excluding both inverted and normal ordering, creating tension with the results of particle physics experiments~\cite{Gariazzo23,Jiang24}.

These bounds are substantially relaxed under a DDE framework modeled by CPL, yielding $\sum m_\nu < 0.11\; \text{eV}$ at 95\% C.L~\cite{DESI:2025zgx}(see also~\cite{Shouvik25}). When expanding the cosmological model to a dynamical dark energy scenario, larger values of $\sum m_\nu$ are allowed by decreasing the value of $w(z)$, since these two parameters are anti-correlated \cite{Hannestad:2005gj,RoyChoudhury:2019hls}. The degeneracy of $\sum m_\nu$ and $w(z)$ is partially broken when CMB, BAO and supernova data are considered concomitantly. 

This work examines how other well-established EoS phenomenological parameterizations, such as Barboza-Alcaniz (BA) and Jassal-Bagla-Padmanabhan (JBP), affect the $\sum m_\nu$ constraints. We also consider a $w(z)$ parameterization that describes the overall behavior of several thawing quintessence models~\cite{deSouza:2025rhv}. In our work, we performed a Markov Chain Monte Carlo (MCMC) analysis combining the latest CMB, SN, and DESI BAO data. We also perform a Bayesian analysis to compare the results of the above parameterizations with respect to the CPL parameterization. In Section 2, we detail the functional forms and the motivation of the DDE models considered. Section 3 presents the methodology adopted in our analysis and the results obtained. In Section 4, we present and discuss our main conclusions.


\section{Dynamical Dark Energy Models} \label{sec:2}

DDE models affect the evolution of the Universe in two primary ways: through the expansion rate and through the cosmological perturbations that give rise to structures. In the case of the Universe's expansion rate, assuming a statistically homogeneous and isotropic Universe that follows the Friedmann-Lemaître-Robertson-Walker (FLRW) metric, we have:
\begin{eqnarray}
H^2 = H_0^2 \left[\rho_r a^{-4}+\rho_m a^{-3}  + \rho_{de}a^{-3}\exp{[\Upsilon(a)]}\right]\;,
\end{eqnarray}
where $w(a)$ represents the EoS as a function of the scale factor and $\Upsilon(a) = \left[\int_1^{a}-3 \left( \frac{w(a)}{a} \right) da\right]$. For the cosmological perturbations the first and most obvious impact of a time-evolving EoS is on the density perturbation and on the fluid velocity. If we assume a Newtonian gauge,
\begin{equation}
    ds^2 = -(1+2\Psi)dt^2 + a^2(t)(1-2\Phi)\delta_{ij}dx^i dx^j,
\end{equation}
then, the Continuity and Euler equations are written as
\begin{subequations}
\begin{equation}
       \dot{\delta} + 3\mathcal{H}(c_s^2 - w)\delta + (1 + w)\left(\theta - 3\dot{\Phi}\right) = 0,
\end{equation}
\begin{equation}
    \dot{\theta} + \mathcal{H}(1-3w)\theta - \frac{k^2 c_s^2 \delta}{1 + w} - k^2\Psi = 0.   
\end{equation}
\end{subequations}

The dark energy also affects the adiabatic sound speed of the fluid $c_s^2$. To investigate the impact of DDE models on the sum of neutrino mass, we will use the following functional forms for the EoS,

\begin{itemize}
  \item The Jassal-Bagla-Padmanabhan parameterization~\cite{Jassal_2005}
  \begin{equation}
        w(a) = w_0+w_a a(1-a)\;.
  \end{equation}
\end{itemize}
The parameterization above consists of a combination of  linear and  quadratic terms in the scale factor. In the current epoch, the term $-w_aa^2$ becomes comparable to $w_aa$, leading to  differences at low-redshifts compared to the CPL model.

\begin{itemize}
    \item The Barboza-Alcaniz parameterization~\cite{Barboza08}    
    \begin{equation}
    w(a)=w_0+w_a\frac{1-a}{a^2+(1-a)^2}\;.
    \end{equation}
\end{itemize}  
This parameterization exhibits a linear behavior at low-redshifts. Unlike the CPL parameterization, which blows up exponentially as the scale factor approaches infinity (for $w_a > 0$), the BA parameterization remains well-behaved at all times, while still allowing for deviations from the CPL scenario. 

\begin{itemize}
    \item The thawing quintessence parameterization~\cite{Carvalho_2006,deSouza:2025rhv}
    \begin{equation}\label{eq6}
    w(a) = -1 + \alpha a^\beta\;.
    \end{equation}
\end{itemize}
Standard quintessence scenarios generally predict an EoS parameter satisfying $w > -1 $, never crossing into the phantom regime. To generalize a class of quintessence potentials, \cite{Carvalho:2006fy} proposed the ansatz,
${1}/{\rho_\phi} ({d\rho_\phi}/{da}) = -{3\alpha}/{a^{1+\beta}}$, 
which results in (\ref{eq6}) when combined with the continuity equation, 
    $\dot\rho_\phi + 3H\rho_\phi[1 + w(a)] = 0$.
For $\alpha > 0$, Eq.~(\ref{eq6}) captures both thawing and freezing behaviors of scalar field models, with $\alpha$ setting the present-day EoS value and $\beta > 0$ controlling the rate of deviation from $w = -1$. When both $\alpha > 0$ and $\beta > 0$, the model reproduces the typical dynamics of thawing quintessence, where the field is initially frozen due to Hubble friction and behaves like a cosmological constant, later evolving to $w(z) > -1$. For a recent analysis of this generalized quintessence scenario in the absence of neutrinos, see~\cite{deSouza:2025rhv}.

\section{Analysis and Results} \label{sec:3}

To obtain constraints on the cosmological parameters for the DDE models above, we perform a Monte Carlo Markov Chain (MCMC) numerical analysis using Planck 2018 temperature and polarization data, including low- and high-multipole measurements of temperature and polarization, combined with lensing (plikHM+TTTEEE+lowl+lowE+lensing)~\cite{planck18}\footnote{It is important to note that DESI uses Planck combined with Data Release 6 of the Atacama Cosmology Telescope (ACT), while in our analysis we use CMB data from the Planck baseline likelihood only.}, along with Pantheon+ catalog SN data~\cite{Brout_2022,Scolnic_2022} and geometric BAO information from the DESI collaboration data release two~\cite{DESI:2025zgx}. For this, we implement the parameterizations in the Boltzmann solver code \textit{Cosmic Linear Anisotropy Solving System} (CLASS)~\cite{lesgourgues11}, in which we interfaced with the \textit{Cobaya} code to perform the MCMC statistical analysis~\cite{
Blas:2011rf,Torrado_2021}. For the purpose of constraining the neutrino mass, we assume the standard degenerate mass spectrum, in which the three massive neutrino species each carry one third of the total mass, i.e., $\sum m_\nu = 3 (m_{\nu,1} = m_{\nu,2} = m_{\nu,3})$. See~\cite{Herold_2025} for a detailed discussion on the choice of neutrino mass hierarchy in cosmological analyses. {In our analysis, we also imposed a prior that the total neutrino mass must satisfy $\sum m_\nu > 0$, with $\sum m_\nu \in [0,5]$. For the EoS parameters our priors are, $w_0 \in [-3,1]$, $w_a \in [-3,2]$, $\alpha \in [0,1]$ and $\beta \in [0,15]$.} 

\begin{figure}[t]
\centering
\includegraphics[width=\columnwidth]{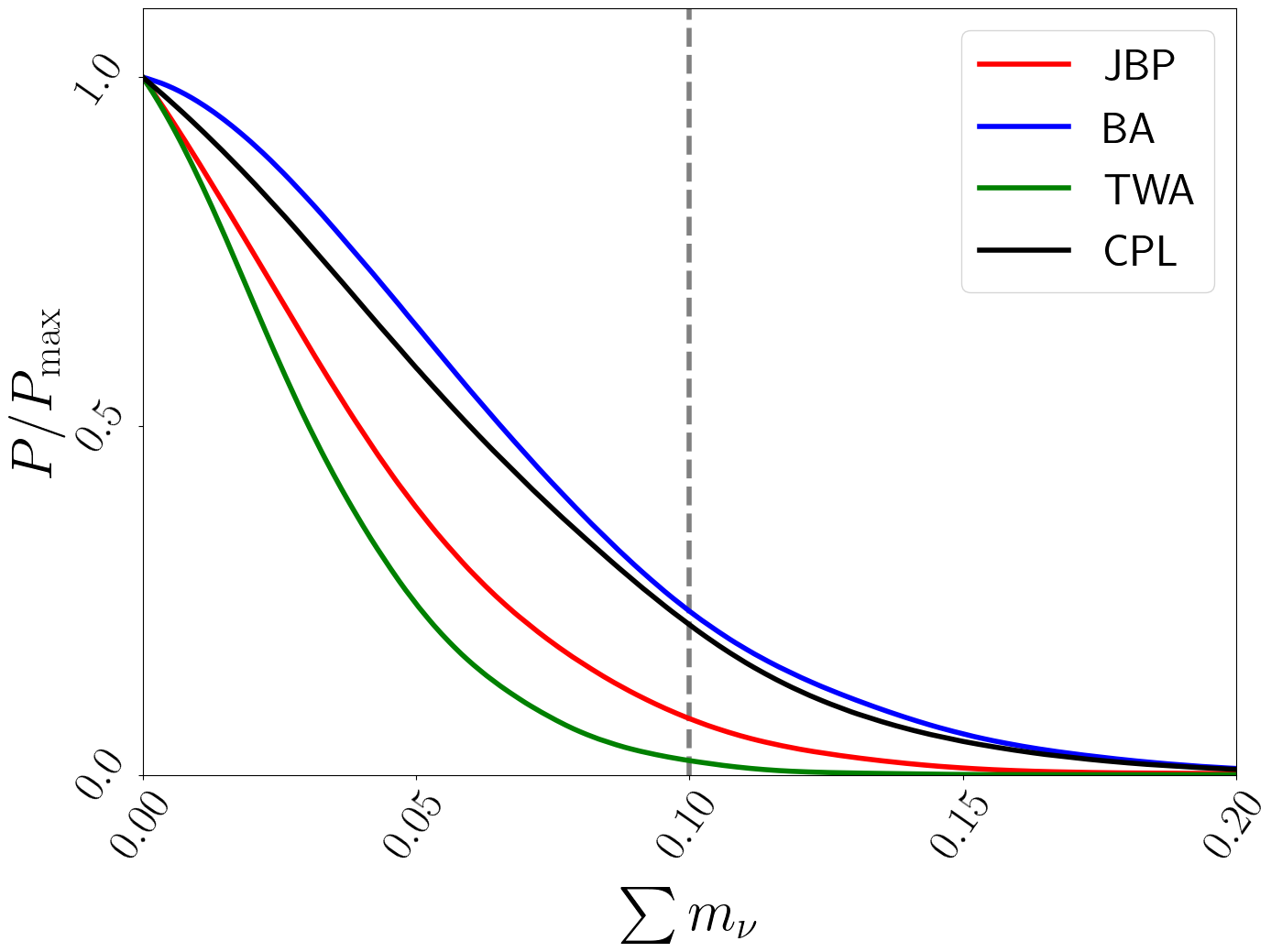}
\caption{Upper bounds on $\sum m_\nu$ from Planck 2018 temperature, polarization and lensing together with Pantheon+ SN and DESI DR2 data assuming JBP, BA, and Thawing parameterizations. The dashed vertical line represents the inverted hierarchy $\sum m_\nu > 0.10$~eV. For comparison, we also show the CPL prediction (black line). }
\label{fig:1}
\end{figure}

Figure~\ref{fig:1} shows the marginalized posterior distribution for $\sum m_\nu$ across the models explored in this work. For comparison, see Fig. 7 of~\cite{elbers2025}, which presents results from the DESI collaboration using DESI DR2 BAO + CMB + Pantheon+ data. In their analysis of the CPL model, the posterior peaks at the prior boundary $\sum m_\nu = 0$, suggesting that the data favor non-physical values of $\sum m_\nu < 0$. This is also true in our analysis: all models show a peak at $\sum m_\nu = 0$. This suggests that regardless of the DDE model, when including DESI data, there is still a preference outward $\sum m_\nu \leq 0$ values. 

\begin{figure*}[t]
\centering
\includegraphics[width=0.85\columnwidth]{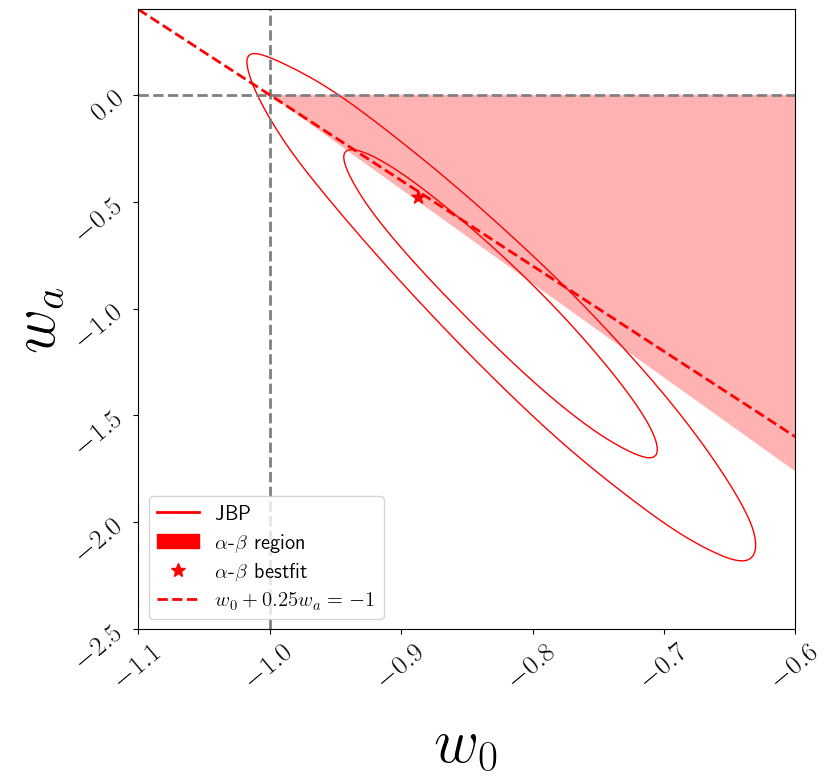}
\includegraphics[width=0.84\columnwidth]{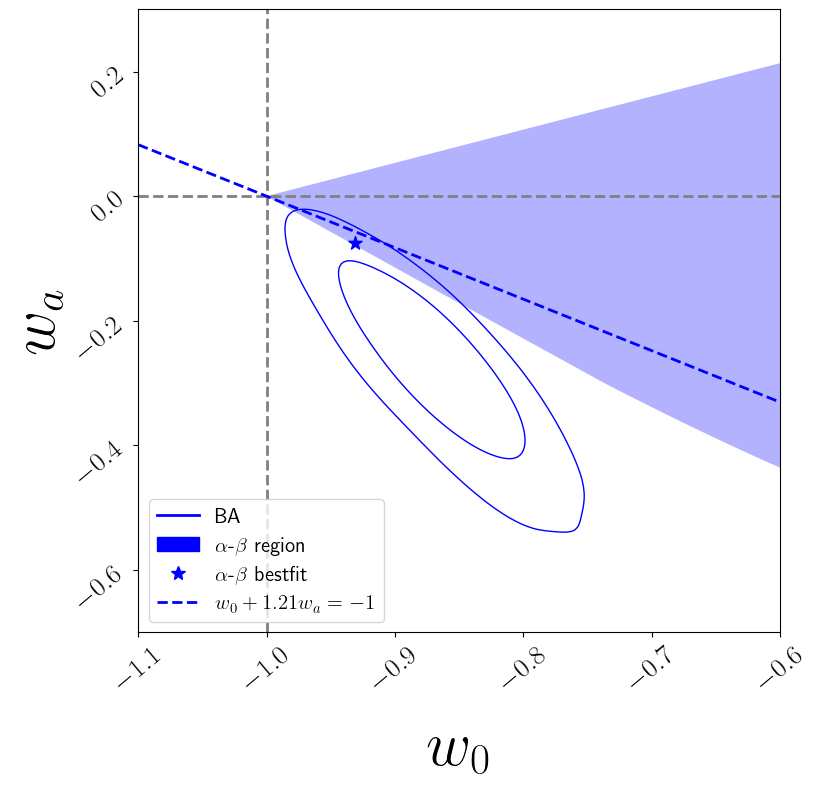}
\caption{$w_0-w_a$ plane for the JBP (upper) and BA (bottom) parameterizations obtained from Planck 2018 temperature, polarization and lensing together with Pantheon+ SN and DESI DR2 BAO data. The dashed line intersection represents the point, $w_a = 0$ and $w_0 = -1$. The shaded red (JBP) and blue (BA) regions represent the full range of $\alpha$–$\beta$ priors projected into the $w_0$–$w_a$ space, while the star markers indicate the projected best-fit $\alpha$–$\beta$ values for each model. The red and blue dashed lines shows the dividing line between quintessence and phantom behavior for each EoS parameterizations. }
\label{fig:2}
\end{figure*}

For the CPL parameterization, the DESI collaboration reports an upper bound of $\sum m_\nu < 0.11$~eV,  a value compatible with both the normal and inverted mass hierarchies. In our analysis, the BA and CPL models give us nearly the same upper bound,  $\sum m_\nu < 0.12$~eV, also consistent with both mass orderings. The JBP, however, provides a tighter constraint, with a 2$\sigma$ upper bound of $\sum m_\nu < 0.096$~eV, indicating a mild preference for the normal hierarchy. Interestingly, the thawing parameterization (\ref{eq6}) imposes the tightest constraint, with $\sum m_\nu < 0.071$~eV. Not only does the thawing model constrain the total neutrino mass at a level comparable to that of the $\Lambda$CDM model, therefore excluding the inverted hierarchy by more than $3\sigma$, while also weakens any preference for a non-physical negative neutrino mass within our data choice~\footnote{We also performed the same analysis considering other SN samples, such as DESY5~\cite{desy5} and Union3~\cite{union3} compilations. Overall, we did not observe significant changes in the above conclusions.}.

In addition, we conducted a Bayesian statistical comparison of all models with respect to the CPL (for more details on this kind of analysis, we refer the reader to \cite{deSouza:2025rhv}). 
Table~\ref{Tab1} presents the 2$\sigma$ upper limits on $\sum m_\nu$ for each model, along with the Bayes factor, ${\Delta\ln{B}}$. 
From this analysis, we find that the preference for both BA and JBP are statistically inconclusive compared to CPL ($|{\Delta\ln{B}}| < 1$) while the TWA model shows a moderate Bayesian preference over CPL, with $|{\Delta\ln{B}}| = 2.66$. 

\begin{table}[]
     \centering
    \resizebox{0.70\columnwidth}{!}{
    \begin{tabular}{|c|c|c|}
        \hline
        \hline
        Model & $\sum m_\nu$ ($2\sigma$) 
        & $\Delta\ln{B}$ \\
        \hline
        CPL  &  $< 0.123$\;eV 
        & $0.0$ \\
        BA   &    $< 0.127$\;eV 
        & $-0.64$ \\
        JBP  &  $< 0.096$\;eV 
        & $-0.60$ \\
        TWA  &  $< 0.0709$\;eV 
        & $2.66$ \\
        \hline
        \hline
    \end{tabular}}
    \caption{$2\sigma$ upper bounds on the sum of neutrino masses $\sum m_\nu$ and the Bayes factor, $\Delta\ln{B}$, for all models discussed in the text. Negative values of $\Delta\ln{B}$ favors the reference model (CPL).} 
    \label{Tab1}
\end{table}

\begin{figure*}[t]
\centering
{\includegraphics[width=0.32\linewidth]{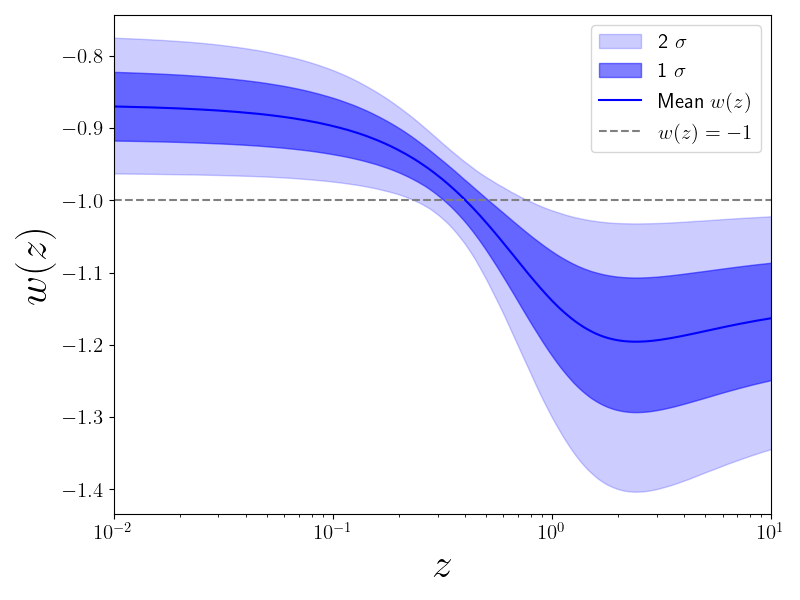}} 
{\includegraphics[width=0.32\linewidth]{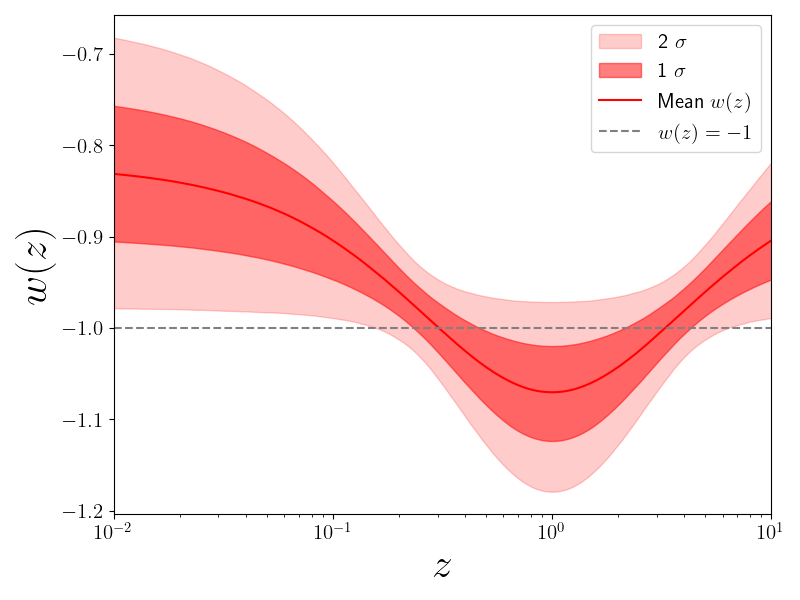}}
{\includegraphics[width=0.32\linewidth]{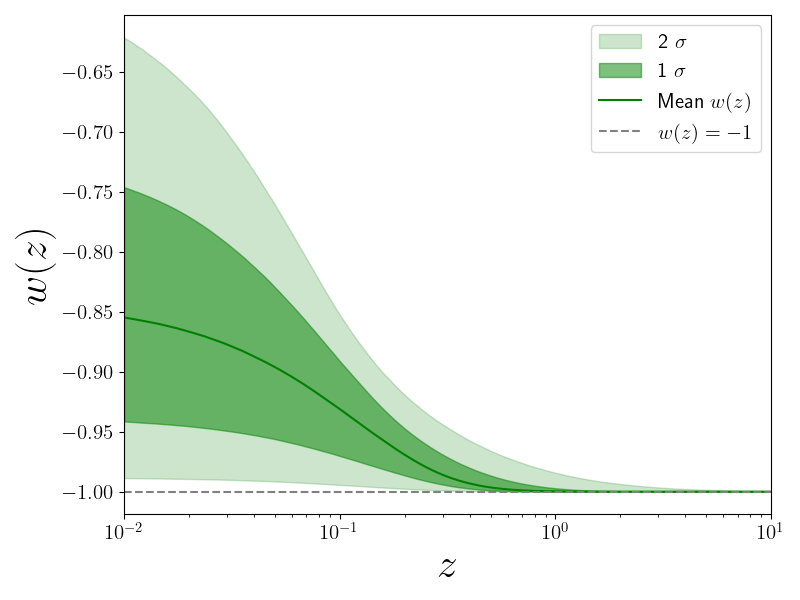}}
\caption{EoS evolution obtained for Planck+DESI+Pantheon+ data assuming the BA (blue), JBP (red) and Thawing (green) parameterizations.}
\label{fig:3}
\end{figure*} 

In Figure~\ref{fig:2}, we show the $w_0$–$w_a$ planes for JBP (upper panel) and BA parameterizations (bottom panel). When treating neutrino mass as a free parameter, the JBP model is consistent with the $\Lambda$CDM scenario within $2\sigma$, while the BA model lies nearly outside the $2\sigma$ boundary. Following the procedure of~\cite{Shlivko:2025fgv}, we also project the $(\alpha, \beta)$ values from the thawing model into the $(w_0, w_a)$ parameter space by matching the expansion history $H(z)$. 
The shaded regions in both panels of Fig.~\ref{fig:2} represent the $\alpha$–$\beta$ values within the full prior range considered in the analysis. This shows that the phantom region in the $w_0$–$w_a$ plane for both models is fully compatible with the scenario of thawing quintessence. The star markers correspond to the best-fit values, $\alpha = 0.15$ and $\beta = 7.84$, projected into the $w_0$–$w_a$ parameter spaces for both models. These best-fit projections lie within the $1\sigma$ and $2\sigma$ confidence regions of the JBP and BA, respectively.

Figure \ref{fig:3} shows the EoS evolution with redshift for each model, using Planck CMB temperature, polarization, and lensing data, along with Pantheon+ SN and DESI BAO. The BA and JBP behave similarly at low redshifts, remaining in the quintessence region, i.e., $w(z) > -1$. The BA model crosses to phantom $w(z) < -1$ at a slightly lower redshift than the JBP, and remains in the phantom region near $w(z) = -1$. Interestingly, for the JBP model, it tends to cross $w(z) = -1$ twice, leaving the quintessence region and remaining very close to $w(z) = -1$ with a trend toward the phantom region until it returns to the quintessential region for high $z$ values. The thawing EoS remains exactly equal to $w(z)= -1$ at high redshifts and begins to evolve toward the $w(z) > -1$ region near $z \approx 0$.


\section{Final Remarks}\label{sec:4}

Our findings in this work highlight several key aspects. First, regardless of the underlying cosmological model, be it $\Lambda$CDM or a DDE scenario, the posterior distribution for $\sum m_\nu$, when DESI data are included, consistently peaks at the lower prior boundary, $\sum m_\nu = 0$. This behavior indicates that the data would prefer values of $\sum m_\nu$ that are negative, an unphysical regime excluded by the prior. Consequently, the derived upper bounds on $\sum m_\nu$ in this context are largely driven by the prior choice. The thawing parameterization provides the strongest constraint on the neutrino mass, $\sum m_\nu < 0.071$ eV at 95\% C.L., comparable to that of $\Lambda$CDM and sufficient to exclude the inverted hierarchy at more than $3\sigma$. The JBP model also shows a mild preference for the normal hierarchy over the inverted, with an upper bound of $\sum m_\nu < 0.096$ eV at 95\% C.L. We find that the JBP model is consistent with $\Lambda$CDM within $2\sigma$ while the BA model sits near the outer limit of $2\sigma$. Both models exhibit a shift toward $\Lambda$CDM when the neutrino mass is included in the analysis as a free parameter. This is significant, as it weakens the apparent preference for DDE models seen in analyses where the total neutrino mass is fixed at $\sum m_\nu = 0.06$ eV.

By mapping the thawing model into the $w_0$–$w_a$ plane, we also demonstrate that this class of quintessence models is fully compatible with the regions occupied by the BA and JBP models, even extending into the phantom regime, which appears to be favored by the data. This suggests that, contrary to the claims of the DESI collaboration of phantom dark energy evidence, the data may in fact be suggesting evidence for quintessence scenarios. {Notably, the thawing model also shows a moderate Bayesian preference over the CPL, with $\Delta\ln{B}= 2.66$.}

Finally, we studied the redshift evolution of the dark energy equation of state. Although they all remain in the quintessence regime at low redshifts, their behaviors diverge at higher redshifts. The thawing model stands out because of its smooth transition from a cosmological constant-like state ($w = -1$) at high redshift to a quintessence-like behavior near the present epoch. Overall, our results provide evidence that thawing quintessence models are not only fully compatible with widely used phenomenological parameterizations when projected in the same parameter space, but also offer tighter bounds on the neutrino mass. The thawing model also shows a strong preference for the normal mass ordering.

\section*{Acknowledgements}
GR and RdS are supported by the Coordena\c{c}\~ao de Aperfei\c{c}oamento de Pessoal de N\'ivel Superior (CAPES). JR thanks the Funda\c{c}\~ao de Amparo \`a Pesquisa do Estado do Rio de Janeiro (FAPERJ) grant No. E-26/200.513/2025. JSA is supported by CNPq grant No. 307683/2022-2 and FAPERJ grant No. 259610 (2021). We also acknowledge the use of the \texttt{Cobaya}, \texttt{CLASS} and \texttt{GetDist}. This work was developed thanks to the computational support of the National Observatory Data Center (CPDON).

\bibliography{bibliography}

\label{lastpage}

\end{document}